\documentstyle[aps,prl]{revtex}

\begin{document}
\draft
\title{Finite-size Scaling and Universality above the Upper Critical
  Dimensionality}
\author{Erik Luijten\cite{email} and Henk W. J. Bl\"ote}
\address{Faculty of Applied Physics, Delft University of Technology, P.O. Box
  5046, 2600 GA Delft, The Netherlands}
\date{\today}
\maketitle
\begin{abstract}
  According to renormalization theory, Ising systems above their upper critical
  dimensionality $d_{\rm u}=4$ have classical critical behavior and the ratio
  of magnetization moments $Q=\langle m^2 \rangle^2 / \langle m^4 \rangle$ has
  the universal value $0.456947\cdots$. However, Monte Carlo simulations of
  $d=5$ Ising models have been reported which yield strikingly different
  results, suggesting that the renormalization scenario is incorrect.  We
  investigate this issue by simulation of a more general model in which $d_{\rm
    u}<4$, and a careful analysis of the corrections to scaling. Our results
  are in a perfect agreement with the renormalization theory and provide an
  explanation of the discrepancy mentioned.
\end{abstract}
% Keywords: finite-size scaling, renormalization,
%           universality, long-range interactions

%\narrowtext

One of the most important contributions to the modern theory of critical
phenomena is Wilson's renormalization theory (see Ref.~\cite{wilson-kogut} for
an early review). This theory explains the existence of a so-called {\em upper
  critical dimensionality\/} $d_{\rm u}$. It predicts that systems with a
dimensionality $d > d_{\rm u}$ exhibit classical exponents and violate
hyperscaling, whereas systems with a lower dimensionality behave
nonclassically.  For Ising-like systems with short-range interactions, $d_{\rm
  u}=4$.  In recent years, a controversy has arisen about the value of the
``renormalized coupling constant'' or ``Binder cumulant''~\cite{cumulant} for
$d>d_{\rm u}$.  On the one hand, a renormalization calculation for hypercubic
systems with periodic boundary conditions~\cite{brez-zj} predicts that the
Binder cumulant assumes a universal value for $d \geq d_{\rm u}$.  On the other
hand Monte Carlo simulations of the five-dimensional Ising
model~\cite{hyper,binder5d,rickw} yielded significantly different results.
Since the renormalization theory forms the basis of our present-day
understanding of phase transitions and critical phenomena, it is of fundamental
interest to examine any discrepancies and inconsistencies with this theory.
Furthermore, there exist several models with a lower value of $d_{\rm
  u}$~\cite{binder5d,lr-rg} where the above-mentioned issue may be of
experimental interest as well.

In this Letter, we answer the question concerning the value of the Binder
cumulant.  One of the key issues is the shift of the ``critical temperature''
in finite systems.  We rederive this shift, which was already calculated in
Ref.~\cite{brez-zj}, from basic renormalization equations and show that the
result agrees with the shifts observed in Refs.~\cite{hyper,binder5d,rickw}.
Furthermore, we determine the Binder cumulant in the context of a more general
Ising-like model with algebraically decaying interactions.  This model is
subject to the same renormalization equations as the aforementioned $d=5$ Ising
model, and effectively reduces to the nearest-neighbor model when the
interactions decay fast enough. For slow decay, the upper critical
dimensionality decreases below 4 and we have thus been able to investigate the
question concerning the universality of the Binder cumulant in the classical
region by means of Monte Carlo simulations of low-dimensional models. This
enabled us to examine a much larger range of system sizes than in the
five-dimensional case.  High statistical accuracies were obtained by using a
novel Monte Carlo algorithm for systems with long-range interactions and we
could resolve various corrections to scaling that are present. The results turn
out to be in complete agreement with the renormalization predictions.

We formulate our analysis in terms of the dimensionless amplitude ratio $Q =
\langle m^2 \rangle^2 / \langle m^4 \rangle$, where $m$ is the magnetization.
This ratio is related to the fourth-order cumulant introduced by
Binder~\cite{cumulant}. In Ref.~\cite{brez-zj}, it is predicted that in
hypercubic short-range Ising-like systems with periodic boundary conditions and
$d \geq 4$, this quantity takes at the critical temperature $T_{\rm c}$ the
universal value $8\pi^2/\Gamma^4(\frac{1}{4}) = 0.456947\cdots$, which is
simply the value of $Q$ in the mean-field model~\cite{ijmpc}. In contrast, the
Monte Carlo simulations in Refs.~\cite{binder5d,rickw} yield the values $Q
\approx 0.50$ and $Q=0.489$~(6), respectively.  In Ref.~\cite{rickw}, this
discrepancy is explained by a size-dependent shift of the ``effective critical
temperature'' $T_{\rm c}(L)$ (defined by, e.g., the maximum in the specific
heat)
\begin{equation}
\label{eq:shift}
  T_{\rm c}(L)=T_{\rm c} - AL^{-d/2} \;,
\end{equation}
which was obtained in Refs.~\cite{hyper,binder5d} from scaling arguments. $L$
denotes the linear system size.

In order to examine this issue we will first outline the theoretical framework
for scaling above $d_{\rm u}$.  As was shown by Br\'ezin~\cite{brezin},
conventional finite-size scaling breaks down for $d \geq d_{\rm u}$.  This is
an example of Fisher's mechanism of {\em dangerous irrelevant variables\/}
(see, e.g., Refs.~\cite{fisher-sa,priv-fish}).  To examine the consequences of
this mechanism for the finite-size scaling behavior, we briefly review the
renormalization transformation for Ising-like models. Near criticality, one can
represent the Hamiltonian for these models by one of the
Landau--Ginzburg--Wilson type,
\begin{equation}
  {\cal H}(\phi)/k_{\rm B}T = \int_{V} {\rm d}^d x \left\{ \frac{1}{2}(\nabla
  \phi)^2 - h\phi + \frac{1}{2} r_0 \phi^2 + u \phi^4 \right\} \;.
\label{eq:hamil}
\end{equation}
$h$ is the magnetic field, $r_0$ is a temperature-like parameter and the term
proportional to $u$ keeps $\phi$ finite when $r_0 \leq 0$.  Under a spatial
rescaling with a factor $b=e^l$ the renormalization equations are, to first
order in $r_0$ and $u$, given in differential form by (see, e.g.,
Ref.~\cite{ma})
\begin{mathletters}
\label{eq:RGdiff}
\begin{eqnarray}
  \frac{{\rm d}r_0}{{\rm d}l} &=& y_{\rm t} r_0 + \alpha u
      \\
  \frac{{\rm d}u}{{\rm d}l} &=& y_{\rm i} u \;,
\end{eqnarray}
\end{mathletters}%
in which $y_{\rm t}$ and $y_{\rm i}$ are the renormalization exponents of the
temperature field and the irrelevant field $u$, respectively, and $\alpha$ is a
constant depending on the dimensionality $d$.  Upon integration, these
equations yield, to first order in $u$,
\begin{mathletters}
\begin{eqnarray}
\label{eq:RG1}
r_0'(b) &=& b^{y_{\rm t}} \left[ (r_0-\tilde{\alpha}u) + \tilde{\alpha}u
  b^{y_{\rm i}-y_{\rm t}} \right] \\
u'(b) &=& b^{y_{\rm i}} u \;,
\end{eqnarray}
\end{mathletters}%
where $\tilde{\alpha}$ is a constant.  This shows that the reduced temperature
$t \equiv (T-T_{\rm c})/T_{\rm c}$ is proportional to $(r_0-\tilde{\alpha}u)$.
Correspondingly, the free energy density $f$ scales as
\begin{equation}
  f \left( t,h,u,\frac{1}{L} \right) = b^{-d}f \left(b^{y_{\rm
      t}}[t+ \tilde{\alpha}ub^{y_{\rm i}-y_{\rm t}}], b^{y_{\rm h}}h, b^{y_{\rm
      i}}u, \frac{b}{L} \right) + g \;,
\label{eq:energy-scale}
\end{equation}
where we have included a finite-size field $L^{-1}$ and $g$ denotes the
analytic part of the transformation. The first term on the RHS can be
abbreviated as $b^{-d}f(t',h',u',b/L)$. For $d \geq 4$, the critical behavior
is determined by the Gaussian fixed point $(t^*,u^*)=(0,0)$. However, for $T
\leq T_{\rm c}$, the free energy is singular at $u=0$.  Hence $u$ is a
dangerous irrelevant variable.  The finite-size scaling properties of
thermodynamic quantities can be obtained by renormalizing the system to size 1,
i.e., setting $b=L$. The number of degrees of freedom then reduces to 1 and the
free energy to
\begin{equation}
  f(t',h',u',1)= \ln \int_{-\infty}^{+\infty} {\rm d}\phi\; \exp \left[
  h' \phi - \frac{1}{2}r_0'(L) \phi^2 - u'(L) \phi^4 \right] \;.
\end{equation}
The substitution $\phi' = \phi/u'^{1/4}$ leads to
\begin{equation}
\label{eq:energy-dang}
  f(t',h',u',1) = \tilde{f}( \tilde{t},\tilde{h} ) \;,
\end{equation}
with $\tilde{t}=t'/u'^{1/2}$ and $\tilde{h}=h'/u'^{1/4}$. Upon renormalization,
the analytic part $g$ of the transformation also contributes to the singular
dependence of the free energy on $t$, see, e.g., Ref.~\cite{ma}. We absorb this
contribution in the function $\tilde{f}$.
Setting $b=L$ and
combining Eqs.\ (\ref{eq:energy-scale}) and~(\ref{eq:energy-dang}) yields
\begin{equation}
\label{eq:energy-rescale}
f\left( t,h,u,\frac{1}{L} \right) = L^{-d}\tilde{f}
\left(L^{y_{\rm t}-y_{\rm i}/2} \frac{1}{u^{1/2}} \left[
  t+\tilde{\alpha}uL^{y_{\rm i}-y_{\rm t}} \right], L^{y_{\rm h}-y_{\rm i}/4}
  \frac{h}{u^{1/4}} \right) \;.
\end{equation}
For $d \geq 4$, $y_{\rm t}=2$, $y_{\rm h}=1+d/2$, and $y_{\rm i}=4-d$. The
first argument on the RHS is the scaled temperature:
\begin{equation}
\label{eq:t-shift}
\tilde{t} = L^{d/2} \frac{1}{\sqrt{u}} \left( t+ \tilde{\alpha}uL^{2-d}
\right) \;.
\end{equation}
Interpreting the term $\tilde{\alpha}uL^{2-d}$ as a shift in the ``effective
critical temperature'' for a finite system, we recover the result of
Ref.~\cite{brez-zj}.

Let us now use the above derivation to examine the shift and rounding of
critical singularities in finite systems.  Observables can be calculated from
the free energy by differentiating with respect to a suitable parameter.
Ignoring the analytic part of the free energy, we can express the thermodynamic
quantities in terms of {\em universal\/} functions of the two arguments that
appear in the RHS of Eq.~(\ref{eq:energy-rescale}).  E.g., the specific heat
can be written as the product of a power of the rescaling factor and a {\em
  universal\/} function of the scaled fields. Let the maximum of this function
occur at $\tilde{t}=c$ ($c$ a constant). Then, the specific heat maximum occurs
at a temperature which differs, in leading orders of $L$, from the critical
temperature by
\begin{equation}
\label{eq:delta-t}
  \Delta t = c\sqrt{u} L^{-d/2} - \tilde{\alpha}u L^{2-d} \;.
\end{equation}

The leading $L$-dependence of Eq.~(\ref{eq:delta-t}) agrees with
Eq.~(\ref{eq:shift}). However, on the basis of Eq.~(\ref{eq:shift}) it is
argued in Refs.~\cite{hyper,binder5d,rickw} that the term between brackets in
Eq.~(\ref{eq:t-shift}) could be replaced by $t+ aL^{-d/2}$, where $a$ is a
nonuniversal constant.  If this argument were correct, it would have serious
consequences for the renormalization scenario: there must be a contribution of
a new type between the square brackets in Eq.~(\ref{eq:RG1}), proportional to
$b^{-d/2}$.  There is no renormalization mechanism known to us which would
yield such a term. Furthermore, in leading orders of $L$ Eq.~(\ref{eq:t-shift})
must be replaced by
\begin{equation}
\label{eq:t-cstshift}
\tilde{t}=L^{d/2} \frac{1}{\sqrt{u}} \left( t+aL^{-d/2} \right) \propto
L^{d/2} t + a \;,
\end{equation}
and in general critical-point values of finite-size scaling functions become
dependent on $a$: they are no longer universal. We illustrate this for the
ratio $Q$. Since the magnetization moments can be expressed in derivatives of
the free energy with respect to the magnetic field, the renormalization theory
predicts
\begin{equation}
\label{eq:qfunc}
Q_L(T)= \tilde{Q}\left( \hat{t} L^{y_{\rm t}^*} \right) + q_1 L^{d-2y_{\rm
    h}^*} + \cdots\;.
\end{equation}
Here $\tilde{Q}$ is a universal function, $\hat{t}$ stands for the argument
between brackets in Eq.~(\ref{eq:t-shift}), and we have introduced the
exponents $y_{\rm t}^* \equiv y_{\rm t}-y_{\rm i}/2$ and $y_{\rm h}^* \equiv
y_{\rm h}-y_{\rm i}/4$.  The additional term $q_1 L^{d-2y_{\rm h}^*}=q_1
L^{-d/2}$ arises from the analytic part of the free energy.  Now suppose that
Eq.~(\ref{eq:t-cstshift}) is correct instead of Eq.~(\ref{eq:t-shift}).  Then
the argument of $\tilde{Q}$ is nonuniversal at the critical point and so
is $Q=\lim_{L \to \infty} Q_L(T_{\rm c})$.  The value calculated in
Ref.~\cite{brez-zj} is then just the particular value of $Q$ for the mean-field
model.

Can we reconcile the renormalization scenario with the Monte Carlo results
obtained up till now?  The evidence for an effective critical temperature as in
Eq.~(\ref{eq:shift}) is based upon the locations of the maxima in the
susceptibility and the specific heat, and those of the inflection points of the
absolute magnetization and the renormalized coupling constant $g_L \equiv -3 +
1/Q_L$.  However, we have seen above that Eq.~(\ref{eq:t-shift}) is fully
compatible with a deviation $\Delta t \simeq L^{-d/2}$ (see
Eq.~(\ref{eq:delta-t})). Therefore, the observed shifts do not provide evidence
for the term proportional to $a$ in Eq.~(\ref{eq:t-cstshift}) and we look for a
different source of the discrepancy between the renormalization and Monte Carlo
results for $Q$.  Equation~(\ref{eq:energy-rescale}) shows that there are
several corrections to scaling which may well account for this.  When
Eq.~(\ref{eq:qfunc}) is expanded in $\hat{t}L^{y_{\rm t}^*}$, the term
proportional to $\tilde{\alpha}$ yields a term $q_2 L^{2-d/2}$.  Furthermore,
when we include a nonlinear contribution in $u$ in (\ref{eq:RGdiff}), factors
$u$ in Eq.~(\ref{eq:energy-rescale}) are replaced by $u(1+\gamma uL^{y_{\rm
    i}})$ and we find an additional term $q_3L^{4-d}$.  Higher powers of these
corrections may also be taken into account in the analysis, as well as the term
$q_1L^{-d/2}$ in (\ref{eq:qfunc}).  However, the determination of these
corrections would require accurate data for a large range of system sizes $L$,
and the high dimensionality of the $d=5$ Ising model presents here a major
obstacle.  The results presented in Refs.~\cite{hyper,binder5d} were based on
$3 \leq L \leq 7$ and therefore the results were by no means conclusive.
Reference~\cite{rickw} used the range $5 \leq L \leq 17$.  Given these limited
ranges of system sizes, it seems uncertain whether all important corrections
have been resolved. Thus the Monte Carlo evidence against the renormalization
result of Ref.~\cite{brez-zj} is not compelling.

Here we follow a different approach to test the renormalization predictions.
In Ref.~\cite{lr-rg}, Fisher, Ma and Nickel investigated the renormalization
behavior of ${\rm O}(n)$ models with ferromagnetic long-range interactions
decaying as $r^{-(d+\sigma)}$ ($\sigma>0$).  The Fourier transform of the
Landau--Ginzburg--Wilson Hamiltonian is quite similar to that of
Eq.~(\ref{eq:hamil}); only the term proportional to $k^2$ is replaced by a term
proportional to $k^{\sigma}$.  Thus the renormalization equations have the same
form as for short-range interactions; only the exponents and the coefficient
$\alpha$ in Eq.~(\ref{eq:RGdiff}) assume different values.  For $0 < \sigma
\leq d/2$ ($d \leq 4$), the Gaussian fixed point is stable and the critical
exponents have fixed, classical, values (and hence hyperscaling is violated).
The upper critical dimensionality is thus $d_{\rm u} = 2\sigma$. In
Fig.~\ref{fig:upperdim}, the regions of classical and nonclassical behavior are
shown as a function of $d$ and $\sigma$. Introducing a parameter $\varepsilon =
2\sigma-d$, we note that the classical exponents apply for $\varepsilon<0$,
just as in the short-range case, where $\varepsilon=4-d$. In the limit $\sigma
\downarrow 0$, each spin interacts equally with every other spin, so that we
can identify this case with the mean-field model.  Thus there is an analogy
between the (short-range) Ising model with $4\leq d < \infty$ and the
long-range model with $0< \sigma \leq d/2$. If the amplitude ratio $Q$ has a
nonuniversal value, we may therefore expect that this manifests itself in the
long-range case as well.

In general, the study of models with long-range interactions is notoriously
difficult, due to the large number of interactions that have to be taken into
account.  However, a novel Monte Carlo algorithm~\cite{ijmpc} of the Wolff
cluster type~\cite{wolff} is available that suppresses critical slowing down
and, in spite of the fact that each spin interacts with every other spin,
consumes a time per spin independent of the system size. Thus we could simulate
models with algebraically decaying interactions in 1, 2, and 3 dimensions and
obtain accuracies that were not feasible up till now (cf.\ Ref.~\cite{monroe}
and references therein).  For $d=1$, the interaction was taken exactly $K
r^{-(d+\sigma)}$, whereas for $d=2$ and $d=3$, the interaction was slightly
modified with irrelevant contributions decaying as higher powers of
$r^{-1}$~\cite{ijmpc}.  To account for the periodic boundary conditions, the
actual spin--spin couplings are equal to the sum over all periodic images.  We
have studied linear system sizes $10\leq L\leq 150000$ for $d=1$, $4\leq L\leq
240$ for $d=2$, and $4\leq L\leq 64$ for $d=3$, generating between one and four
million Wolff clusters per simulation.  The ranges of system sizes are larger
than in Refs.~\cite{hyper,binder5d,rickw}, and more intermediate values of $L$
are available.  These facts, as well as the high statistical accuracy of the
Monte Carlo results, allowed us to resolve the leading finite-size corrections
in the~$Q_L$.

The finite-size scaling analysis was based on the Taylor expansion of the
renormalization prediction for $Q_L$ near criticality:
\begin{equation}
\label{eq:qscal}
Q_L(T)= Q + p_1 \hat{t} L^{y_{\rm t}^*} + p_2 \hat{t}^2 L^{2y_{\rm t}^*} +
p_3 \hat{t}^3 L^{3y_{\rm t}^*} + \cdots + q_1 L^{d-2y_{\rm h}^*} + \cdots
+ q_3 L^{y_{\rm i}} + \cdots \;.
\end{equation}
The coefficients $p_i$ and $q_i$ are nonuniversal and the renormalization
exponents are $y_{\rm t}=\sigma$, $y_{\rm h}=(d+\sigma)/2$, and $y_{\rm
  i}=2\sigma-d$.  The corresponding values $y_{\rm t}^*=d/2$ and $y_{\rm
  h}^*=3d/4$ coincide with those in the short-range case. In addition to the
corrections to scaling in Eq.~(\ref{eq:qscal}) we have also included higher
powers of $q_3 L^{y_{\rm i}}$, which become important especially when $\sigma$
is close to $d/2$. In fact, omitting these corrections yielded estimates for
$Q$ close to those obtained in Refs.~\cite{binder5d,rickw}, although the
residuals strongly indicated the presence of additional corrections. This
confirms the assumption that the discrepancy between the $d=5$ Monte Carlo
results and the renormalization calculation is caused by corrections to
scaling. Furthermore, the coefficient $\tilde{\alpha}$ in
Eq.~(\ref{eq:t-shift}) is very small in all cases, in accordance with the fact
that this correction term could not be resolved in Ref.~\cite{rickw}.  An
extensive analysis of the data will be presented elsewhere. We have fixed all
exponents at the theoretical values, in order to minimize the uncertainty in
$Q$. The results presented in Table~\ref{tab:qkc} show that the agreement
between the renormalization prediction for $Q$ and the Monte Carlo data is
excellent.

It could, for the purpose of comparison, be of some interest to make a
correspondence between systems with short-range interactions in $d>4$
dimensions and $d'$-dimensional systems with long-range interactions decaying
as $r^{-(d'+\sigma)}$. Such a correspondence is possible by expressing the
various finite-size scaling relations in terms of the number of particles $N$
instead of the linear system size $L$. Then the dependence of the thermal and
magnetic exponents on the dimensionality is adsorbed in the parameter $N=L^d$
(or $L^{d'}$) and the renormalization predictions for both models differ only
in the (modified) irrelevant exponents, $(4-d)/d$ and $(2\sigma-d')/d'$,
respectively. For both models, these exponents vary between $0$ and $-1$ in the
classical range, and the matching condition appears as
$\frac{\sigma}{d'}=\frac{2}{d}$. Hence, we may compare the $d=5$ (short-range)
Ising model with the $\sigma=\frac{2}{5}d'$ long-range model, i.e.,
$\sigma=0.4$, $0.8$, and $1.2$ for $d'=1$, $2$, and $3$, respectively. In this
sense the present work approaches the nonclassical regime even closer than
Refs.~\cite{hyper,binder5d,rickw}.

Finally, we remark that models with long-range interactions provide an
effective way to explore scaling properties above the upper critical
dimensionality. E.g., the approach adopted in this Letter may be generalized to
planar, Heisenberg and $q$-state Potts models, including percolation problems.
For $\sigma < 2$, $d_{\rm u}$ is reduced by a factor $\sigma / 2$ in the case
of long-range interactions.


\begin{figure}
\caption{Dimensionality versus decay parameter $\sigma$ for various models.
  Short-range models are described by $\sigma=2$.  The open circles indicate
  the models investigated in this article and the black circle marks that
  of Refs.~\protect\cite{hyper,binder5d,rickw}.}
\label{fig:upperdim}
\end{figure}


\begin{table}
\caption{The ratio $Q$ and critical coupling $K_{\rm c}$ for systems with
  long-range interactions in 1, 2, and 3 dimensions, for several values of the
  parameter $\sigma$ in the range $0 < \sigma \leq d/2$.  The numbers between
  parentheses represent the errors in the last decimal places.}
\label{tab:qkc}
\begin{tabular}{c|d|d|d}
$d$ & $\sigma$ & $Q$         & $K_{\rm c}$ \\ \hline
1   & 0.1      & 0.4584 (14) & 0.047618 (2)  \\
1   & 0.2      & 0.4573 (28) & 0.092234 (5)  \\
1   & 0.25     & 0.4564 (22) & 0.114137 (6)  \\
1   & 0.3      & 0.4590 (45) & 0.136106 (9)  \\
1   & 0.4      & 0.4569 (34) & 0.181150 (10) \\ \hline
2   & 0.2      & 0.4573 (10) & 0.028533 (3)  \\
2   & 0.4      & 0.4565 (17) & 0.051824 (4)  \\
2   & 0.6      & 0.4546 (52) & 0.071358 (8)  \\
2   & 0.8      & 0.4570 (55) & 0.088089 (7)  \\ \hline
3   & 0.2      & 0.4557 (18) & 0.0144344 (14) \\
3   & 0.4      & 0.45686 (8) & 0.0262927 (15) \\
3   & 0.6      & 0.4554 (17) & 0.036045 (3)   \\
3   & 0.8      & 0.4562 (13) & 0.044034 (2)   \\
3   & 1.0      & 0.4580 (25) & 0.050517 (3)   \\
3   & 1.2      & 0.4556 (26) & 0.055678 (2)   \\
3   & 1.4      & 0.460 (9)   & 0.059669 (3)   \\
\end{tabular}
\end{table}


\begin{references}
\bibitem[*]{email} E-mail: {\tt erik@tntnhb3.tn.tudelft.nl}
\bibitem{wilson-kogut} K.G. Wilson and J. Kogut, Phys.\ Rep.\ {\bf 12C}, 75
  (1974).
\bibitem{cumulant} K. Binder, Z. Phys.\ B {\bf 43}, 119 (1981).
\bibitem{brez-zj} E. Br\'ezin and J. Zinn-Justin, Nucl.\ Phys.\ B {\bf 257}
  [FS14], 867 (1985).
\bibitem{hyper} K. Binder, M. Nauenberg, V. Privman and A.P. Young, Phys.\
  Rev.\ B {\bf 31}, 1498 (1985).
\bibitem{binder5d} K.  Binder, Z. Phys.\ B {\bf 61}, 13 (1985).
\bibitem{rickw} C. Rickwardt, P. Nielaba and K. Binder, Ann.\ Phys.\ (Leipzig)
  {\bf 3}, 483 (1994).
\bibitem{lr-rg} M.E. Fisher, S.-k. Ma and B.G. Nickel, Phys.\ Rev.\ Lett.\ {\bf
    29}, 917 (1972).
\bibitem{ijmpc} E. Luijten and H.W.J. Bl\"ote, Int.\ J. Mod.\ Phys.\ C {\bf 6},
  359 (1995).
\bibitem{brezin} E. Br\'ezin, J. Physique {\bf 43}, 15 (1982).
\bibitem{fisher-sa} M.E. Fisher, in {\it Proceedings of the Summer School on
    Critical Phenomena, Stellenbosch, South Africa, 1982}, edited by F.J.W.
  Hahne (Springer, Berlin, 1983).
\bibitem{priv-fish} V.  Privman and M.E. Fisher, J. Stat.\ Phys.\ {\bf 33}, 385
  (1983).
\bibitem{ma} S.-k. Ma, {\it Modern Theory of Critical Phenomena\/}
  (Addison--Wesley, Redwood, California, 1976).
\bibitem{wolff} U. Wolff, Phys.\ Rev.\ Lett.\ {\bf 62}, 361 (1989).
\bibitem{monroe} J.L. Monroe, J.\ Stat.\ Phys.\ {\bf 76}, 1505 (1994).
\end{references}
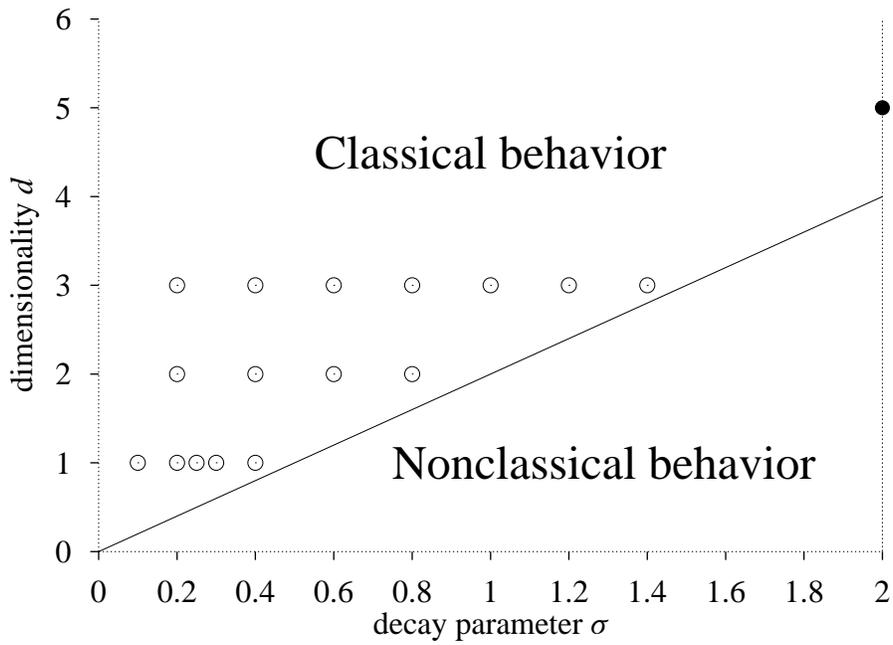
\end{document}